\documentclass[a4paper,11pt]{article}
\pdfoutput=1

\usepackage[T1]{fontenc} 

\usepackage{amssymb,amsmath,amsfonts,graphicx,color}

\usepackage{jheppub}

\title{Black hole essay}

\author{Enrico Brehm}
\affiliation{Max Planck Institute for Gravitational Physics\\ Albert Einstein Institute \\ Potsdam-Golm, D-14476, Germany}
\emailAdd{brehm@aei.mpg.de}

\abstract{This essay gives a very general introduction to Schwarzschild black holes. First, it focuses on some of its classical features as solutions to Einstein's theory of gravity. In the second part it discusses briefly some specific quantum aspects and how a black hole processes quantum information. No previous knowledge about black holes, gravity or quantum mechanics is required.}

\begin{document}
\maketitle

\section{The classical viewpoint}

Black holes are some of the strangest phenomenons in our univers. Here, we first want to discuss them as objects of a classical physical theory, where \textit{classical} means that we forget about all possible quantum effects and their consequences. In the present case the underlying classical theory is Einstein's theory of gravity \cite{Einstein}. It describes space and time in terms of a collection of fields 
whose behaviour is dictated by the Einstein equations. 

One natural question 
to ask is how does this theory describe space and time around a massive spherical object like a star? The solution to this was found by Schwarzschild \cite{Schwarzschild}. It is valid all around any static round object and, remarkably, only depends on its mass. However, quite strange things can happen when all that mass is packed within a specific radius named after Schwarzschild. Then a so-called event horizon forms at the Schwarzschild radius and we start talking about a black hole. 

Before we shed more light on the strangeness of black holes, let us try to get some intuition for the circumstances under which black holes can appear. The Schwarzschild radius $r_s$ and the mass $M$ of a spherical object have a very easy relation: they are proportional to each other, $r_s = a\cdot M$, with $a$ being very small when we measure in standard units. Let us for example consider an object with the mass of earth, then its Schwarzschild radius is given by only about 9 mm! No physical process is known in which all of earth can be compressed that much and it seams quite unlikely that there are many (if at all) black holes with the mass of earth. The situation changes when we consider bigger and bigger masses. This is because the volume within the Schwarzschild radius, \textit{i.e.}, the space where we can store all the mass to form a black hole, increases much faster. In fact, 
if we double the mass we get roughly eight times more volume to store it. The formation of black holes becames easier the heavier they are. It is known that there are mechanisms at the end of the lifetime of very massive stars that lead to the formation of so-called stellar black holes \cite{Oppenheimer}. Even heavier black holes can for example develop when stellar ones merge. 

Let us return to the promised strangeness of black holes. Actually, if we are far away from the black hole, it is not much different from any other stellar object of the same mass. The only big difference is that we do not see any light originating from the black hole. An interesting effect, that occurs when we come closer to it, is that time for us passes slower than for those who stay away from the black hole. However, the effect is not tractable in a direct way. Any clock that we carry with us behaves completely fine from our perspective. Only if we return back to a place far away from the black hole and compare the times passed we could see a difference. This effect appears, in fact, for any massive object that we come close to and is not only a feature of black holes. Remember that the solution to Einstein's equations, which in particular tells us how time behaves, outside the spherical object only depends on its mass! However, for any stellar object that is not a black hole we would at some point reach that object and enter it. 
Inside of it the solutions do depend on its specifics. One can show that effects like time dilation can not increase further arbitrarily. If we, however, come closer and closer to the black hole the effect will grow without a bound until we reach the previously mentioned event horizon. 

Passing the event horizon has some severe consequences. If we take for example the latter observation of unbounded time dilatation serious, we come to the conclusion that in the moment that we spend on the horizon \textit{all} time passes for anything outside the black hole. The end of all things happens in the rest of the universe, and in fact after the moment in which we enter the black hole, there is no way back. This is also visible in a remarkable and strange feature of Schwarzschild's solution. If we compare it inside and outside of the event horizon one observes that global time and the radial direction interchanged their meaning. In a world outside the black hole everything and everyone has to move forward in time. This is a fundamental feature of Einstein's theory. Inside the event horizon the radial direction takes the place of time with the drastic consequence that, no matter what, we have to move forward in this direction, where forward means towards the center of the black hole. There really is no way back! Not even the most powerful rocket that we could imagine can prevent us from finally reaching the very center of the black hole, where gravitational forces become immeasurably strong and at the latest there the quantum features of black holes and gravity itself must show their face. 

However, other quantum aspects of black holes will be visible much earlier. Some of them will be discussed in what follows. 

\section{Contact with the quantum world and information processing}

We have seen before that Schwarzschild's solution only depends on the mass of the object. But what happens with all the information about the object that collapsed into a black hole? It persisted of many different particles, it had a temperature, a matter distribution, a specific spectrum of radiation, and so on. If we only believe in the classical world, then all that information is hidden behind the event horizon after the formation of the black hole. Then it is by no means tractable for anyone outside the black hole. In a classical world this is not a big problem. It is at most sad that no one outside the black hole can get the information but causes no issues concerning consistency of the theory. 

However, we know that our world is in fact not a classical one, and our knowledge about the quantum theories that describe at least ordinary matter in our universe is rather decent. Using this knowledge Hawking could show that quantum effects near the event horizon of a black hole lead to a constant flow of particles away from it \cite{Hawking}. A black hole radiates and, hence, must loose mass over time. If we wait long enough a black hole evaporates either completely or until some tiny remnant of it is left. 

Where is all the information after the evaporation? Now that we include some quantumness in the description of black holes this question becomes very important. Thoughtless processing of quantum information can easily lead to inconsistencies. For example, information has to be spread very fast inside the black hole. Otherwise it would be possible to copy quantum states which is strictly forbidden in any consistent quantum theory. To be honest there is no real consensus among the physics community on how the black hole deals with quantum information. One possibility might be that it is hidden in its Hawking radiation. If we wait long enough and collect a sufficient amount of it we might be able to regain all the information we want. However, there are many more gedankenexperiments concerning similar issues. Finding a convincing and self-consistent description of black holes in contact with the quantum world will probably be an important step towards a quantum description of gravity itself. This might be one of the next big steps in theoretical physics!

At last, let us try to get some intuition on how important quantum effects of a black hole are. If we first consider an ordinary stellar black hole of, say, four times the mass of our sun, then its Hawking radiation can be associated with a temperature only roughly a hundred millionth Kelvin\footnote{Kelvin is the standard measure of temperature in physics. A change of one Kelvin in temperature is the same as the change of one $^\circ$C. } above the absolute zero temperature. It, hence, plays almost no role in describing the everyday physics of that black hole. This is true for any stellar (or even heavier) black hole. Next let us consider a coin of, say, five gram. Quantum effects of that coin do not play any significant role in its everyday physics. It can be described almost perfectly by classical theories. However, if we consider a black hole of the same mass, things look rather different. As a 
(partially) quantum object it radiates and evaporates within a tiny fraction of a second. All its mass converts into energy which results in an explosion three times stronger than the bomb dropped on Hiroshima. We see that for black holes quantum effects play a significant role much earlier than for matter under ordinary circumstances.


\begin{thebibliography}{1}

\bibitem{Einstein}
A.~{Einstein}.
\newblock {Die Grundlage der allgemeinen Relativit{\"a}tstheorie}.
\newblock {\em Annalen der Physik}, 354:769--822, 1916.

\bibitem{Hawking}
S.~W. Hawking.
\newblock Particle creation by black holes.
\newblock {\em Communications in Mathematical Physics}, 43(3):199--220, Aug
  1975.

\bibitem{Oppenheimer}
J.~R. {Oppenheimer} and G.~M. {Volkoff}.
\newblock {On Massive Neutron Cores}.
\newblock {\em Physical Review}, 55:374--381, February 1939.

\bibitem{Schwarzschild}
K.~{Schwarzschild}.
\newblock {On the Gravitational Field of a Mass Point According to Einstein's
  Theory}.
\newblock {\em Abh.~Konigl.~Preuss.~Akad.~Wissenschaften Jahre 1906,92,
  Berlin,1907}, 1916, 1916.

\end{thebibliography}
\end{document}